\documentclass{mem}
\usepackage{natbib}\usepackage{txfonts}\usepackage{balance}
\usepackage{graphicx}
\usepackage{txfonts}
\usepackage[a4paper]{hyperref}
\idline{79}{3}
\begin{document}
\def\teff{$T\rm_{eff }$}
\def\kms{$\mathrm {km s}^{-1}$}

\title{
Metallicity estimates of Galactic Cepheids based on Walraven photometry
}

   \subtitle{}

\author{
S. Pedicelli,\inst{1,2} 
J. Lub,\inst{3}
J.W. Pel,\inst{4} 
B. Lemasle,\inst{5,6}    
G. Bono,\inst{1,7}
P. Fran\c{c}ois,\inst{6}
D. Laney,\inst{8}
A. Piersimoni,\inst{9}
F. Primas,\inst{1}
M. Romaniello,\inst{1} 
R. Buonanno,\inst{2}
F. Caputo,\inst{7}
S. Cassisi,\inst{9}
F. Castelli,\inst{10}
A. Pietrinferni,\inst{9}
J. Pritchard\inst{1}\\
}

\offprints{S. Pedicelli}
 
\institute{
European Organization for Astronomical Research in the Southern Hemisphere (ESO)
Karl-Schwarzschild-Strasse 2, D-85748 Garching bei Munchen
\email{spedicel@eso.org}
\and
Universita' di Roma "Tor Vergata"
Via Orazio Raimondo,18 00173 Roma, Italy 
\and
Leiden Observatory NL-2300 RA  Leiden
The Netherlands
\and
Kapteyn Institute, University of Groningen,
PO Box 800, 9700AV Groningen,
The Netherlands
\and
Universit\'e de Picardie Jules Verne, Facult\'e des Sciences, 33 Rue Saint-Leu, 80039 Amiens Cedex 1, France
\and
Observatoire de Paris-Meudon, GEPI, 92195 Meudon Cedex, France
\and
INAF Osservatorio Astronomico di Roma, Via Frascati 33,
Monte Porzio Catone, Italy 
\and
South African Astronomical Observatory, 
PO Box 9, 7935 Observatory, South Africa
\and
INAF, Osservatorio Astronomico di Collurania, 
64100 Teramo, Italy
\and
INAF Osservatorio Astronomico di Trieste, 
Via Tiepolo 11, 34143 Trieste, Italy 
}

\authorrunning{Pedicelli}

\titlerunning{Galactic Cepheid metallicities based on Walraven photometry}

\abstract{
We present new empirical and theoretical calibrations of two photometric 
metallicity indices based on Walraven photometry. The empirical calibration
relies on a sample of 48 Cepheids for which iron abundances based on 
high resolution spectra are available in the literature. They cover a broad 
range in metal abundance (-0.5 $\le$[Fe/H]$\le$+0.5) and the intrinsic 
accuracy of the Metallicity Index Color (MIC) relations is better than 
0.2 dex. The theoretical calibration relies on a homogeneous set of 
scaled-solar evolutionary tracks for intermediate-mass stars and on 
pulsation predictions concerning the topology of the instability strip.  
The metal content of the adopted evolutionary tracks ranges from $Z=0.001$ to 
$Z=0.03$ and the intrinsic accuracy of the MIC relations is better than 0.1 dex.  
}
\maketitle{}

\section{Introduction}
Classical Cepheids present several advantageous features when compared with 
field giant stars. These are illustrated by the points below: 
{\em Distance indicators --} they obey to well defined optical and near-infrared 
(NIR) Period-Luminosity (PL) relations and their distances can be estimated with an 
accuracy of the order of a few percent (Benedict et al. 2006; Marconi et al. 2006;
Natale et al. 2007; Fouque et al. 2007). They are also the most popular 
primary distance indicators to calibrate secondary indicators such as the 
SN type Ia \citep{ri05}.   
{\em Stellar tracers --} they are distributed across the Galactic disk, and 
therefore, they can be adopted to trace the radial distribution of intermediate-age
stars \citep{k63}. In particular, they provide robust constraints on the iron and 
heavy elements radial gradients \citep[][and references therein]{le07}.   
{\em Bright stars--} they are bright giants that can be easily identified due to 
their intrinsic variability and for which accurate photometric and spectroscopic 
data can be collected.
However, Classical Cepheids also present a few drawbacks.
{\em Short lifetime --} the lifetime they spend inside the instability strip is 
approximately two orders of magnitude shorter than the central hydrogen burning
\citep{bo00}. This together with the low spatial density and the high 
reddening account for the limited 
sample of Galactic Cepheids currently known \citep{f95}.  
{\em Time series --} Accurate mean magnitudes and colors require a detailed 
time sampling along the pulsation cycle. The Cepheid periods range from a few 
days to hundred days. Therefore, well sampled multiband light curves require 
long observing runs.   
In this investigation we decide to take advantage of the Walraven photometry 
for 174 Galactic Cepheids collected by Pel (1976) and by Lub \& Pel (1977).  
The Walraven VBLUW photometric system was originally designed to 
derive the intrinsic properties of early-type stars. The bands were 
selected to measure the features of the hydrogen spectrum, and three of the five 
bands are located in the ultraviolet spectral regions ($\lambda_V=$5405 \AA, 
$\lambda_B=$4280 \AA, $\lambda_L=$3825 \AA, $\lambda_U=$3630 \AA, 
$\lambda_W=$3240 \AA). The reader interested in a more detailed description 
concerning the ingenious crystal optics filter and the instrumentation is 
referred to Walraven \& Walraven (1960), 
Lub \& Pel (1977), and to Pel \& Lub (2007). Because of its sensitivity to the 
Balmer jump and the slopes of the Balmer and Paschen continua, the Walraven system 
turned out to be also very useful for determining 
the intrinsic parameters of A, F, and G-type stars.
For this reason, it has been used for the study of Galactic and Magellanic Cepheids 
\citep{p76}; \citep{p81} and for field RR Lyrae stars \citep{l77}. 
Observations with the Walraven five-channel photometer, attached to the 91-cm 
`Lightcollector' reflector, started in 1958 at the Leiden Southern Station
in Broederstroom, South-Africa. After 20 years in South-Africa the telescope 
and the photometer were moved to ESO/La Silla in Chile. The observations in the 
new site started in March 1979 and continued for a dozen years until the 
decommissioning of the photometer in 1991. 
Thanks to the better photometric conditions of La Silla, this last phase of 
the operational life of the instrument was particularly fruitful.
During 32 years of almost uninterrupted observations, a very large amount of 
high-quality data was obtained for many types of stars, but the emphasis was 
on OB-associations, clusters, Galactic and Magellanic Cepheids ($\sim$ 200), 
Galactic RR Lyrae stars ($\sim$ 100) and faint F/G stars in the disk and in 
the inner Galactic halo.

\begin{figure}
\centering
\resizebox{\hsize}{!}{\includegraphics{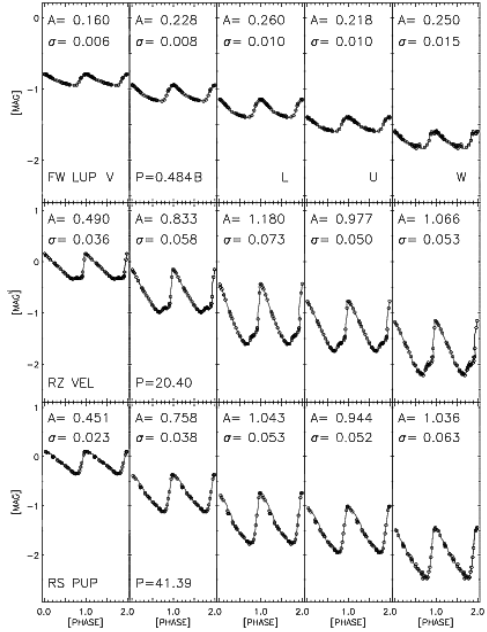}}
\caption{From left to right light curves in the Walraven V,B,L,U,W bands 
for three Cepheids. From top to bottom short, intermediate, and long period 
Cepheids (see labeled values). For each band are also plotted the intrinsic 
scatter of the fit with a cubic spline and the amplitude.}
\label{fig1}
\end{figure}

\begin{figure}
\centering
\resizebox{\hsize}{!}{\includegraphics{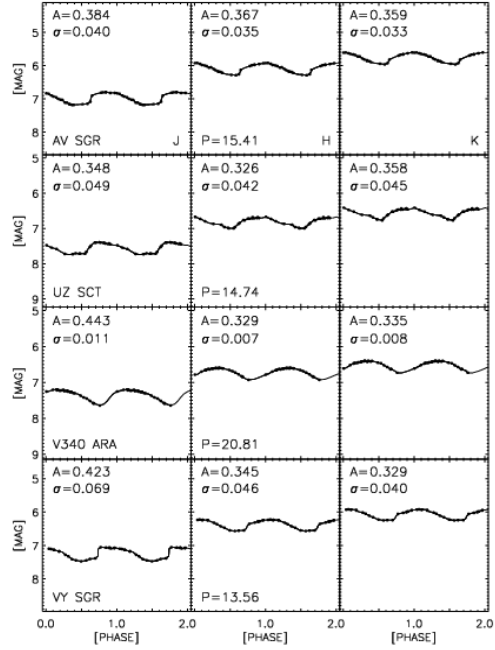}}
\caption{
From left to right J,H,K-band light curves for four Galactic Cepheids. 
The symbols are the same as in Fig. 1. 
}
\label{fig2}
\end{figure}

\section{Data and Theory} 
 
The current Cepheid sample was selected from the observing runs 1962 \citep{w64} and 
1970-1971 \citep{p76} at the Leiden Southern Station (South Africa). 
The sample includes 174 Galactic Cepheids and it is 
82\% complete for all known Cepheids brighter than $V=11.0$ mag at minimum
and south of declination +15$^{\circ}$. Fig. 1 shows the light curves 
in the five bands for three Cepheids of our sample, namely FW LUP, RZ VEL and RS PUP.  
Data plotted in this figure show that individual simultaneous VBLUW measurements 
present an accuracy of the order of few millimag. For each object have been 
collected at least 30 phase points that properly cover the entire pulsation 
cycle. This means that the intrinsic accuracy of the mean magnitudes estimated 
by fitting a cubic spline is of the order of a few hundredths of magnitude.\\
In order to constrain the possible occurrence of systematic errors in the optical 
photometry we are also collecting accurate multiband $J,H,K$ NIR data. At present, 
accurate NIR data are available for 65 Cepheids in the Walraven sample. 
For these data the uncertainties of each phase point range from $0.005$ to $0.007$ for 
$K < 6$ mag, deteriorating to about $0.012$ at $K=8.6$. This implies an accuracy 
in the mean magnitudes of about $0.002-0.005$, depending on the number of points. 
However, the dominant uncertainty on the mean magnitudes is probably due to the 
absolute calibration which is $0.01$.  
For the other Cepheids we use the mean NIR magnitudes from the 2MASS catalog 
\citep{cu03}. Fig. 2 shows the J, H and K-band light curves for four 
Cepheids in our sample.
In order to avoid systematic uncertainties in the metallicity estimate we 
adopted the reddenings given in the Catalog of Classical Cepheids 
\citep{f95} and removed the objects that are members of binary systems 
\citep{sza03}. We ended up with a sample of 151 Cepheids for which we 
searched in the literature for iron measurements based on high resolution 
spectra. We found accurate iron abundances for 48 Cepheids   
(Andrievsky et al. 2002a,b,c; Andrievsky et al. 2004; Luck et at. 2006; 
Mottini et al. 2006; Lemasle et at. 2007). Fortunately enough, the 
calibrating Cepheids cover a broad range in metallicity, namely 
-0.5 $\le$[Fe/H]$\le$0.5. 

\subsection{Interstellar reddening correction}
Before any physical information can be derived from photometric 
data, all colors and magnitudes have to be corrected for interstellar 
reddening. In order to constrain the absorption coefficients, detailed 
observations of a sizable sample of standard stars in the southern 
hemisphere have been performed during three different periods 
(1960-61, 1970-71, 1980-81). There are very small differences between these 
three absolute calibrations of the Walraven system and transformations 
are very well studied. 
The zeropoint was set by the measurements of one B star passing through
the zenith in South Africa ($\omega^1$ Sco, spectral type B1.5V,
$V_J=3.96$ and $(B-V)_J=-0.04$, HD 144470). We underline that we used the 
data from South Africa (1970-71) transformed into the photometric system
as valid for La Silla (1979-1991) and then we were able to apply the relations 
between Johnson and Walraven systems. 
On the basis of the quoted data the following relations were derived by Lub 
and Pel between magnitudes, colors and color excesses in the Johnson and in the Walraven
systems:
{\small\noindent 
$V_J=6.886-2.5V-0.082(V-B)$\\
$(B-V)_J=2.57(V-B)-1.02(V-B)^{2}+0.05(V-B)^{3}$\\
$E(B-V)_J/E(V-B)=2.375-0.169(V-B)$\\
$A_{V_J}/E(B-V)=3.17-0.16(V-B)-0.12E(V-B)$\\
}
while the ratios between different color excess in the Walraven system are the following: 
{\small\noindent 
$E(B-U)/E(V-B)=0.61$\\
$E(U-W)/E(V-B)=0.45$\\
$E(B-L)/E(V-B)=0.39$\\
}
Thanks to the fact that the Walraven photometric system includes 
five bands, we can also define three reddening-free color indices.  
For a standard interstellar extinction law \citep{ca89} 
they are the following:
{\small\noindent 
$[B-U]=(B-U)-0.61(V-B)$\\
$[U-W]=(U-W)-0.45(V-B)$\\ 
$[B-L]=(B-L)-0.39(V-B)$\\
} 
It should be noted that, the Walraven system uses units of $log_{10}(Intensity)$
instead of magnitudes $(-2.5*log_{10}(Intensity))$, therefore
for transforming the distance moduli based on Johnson photometry 
we adopt $DM_W= -0.4* DM_J$.

\begin{figure}
\centering
\resizebox{\hsize}{!}{\includegraphics{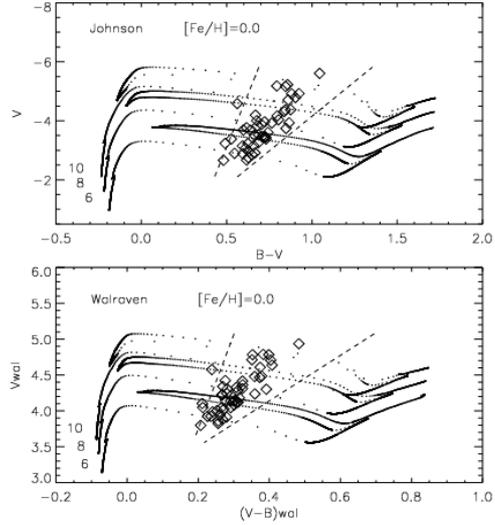}}
\caption{Top: $V_J$,$B_J-V_J$ Johnson Color-Magnitude Diagram for calibrating 
Cepheids (diamonds). Solid lines display evolutionary tracks at solar 
chemical composition (Y=0.273, Z=0.0198) and different stellar masses 
(see labeled values). Individual distances for calibrating Cepheids 
were estimated using the NIR PL relations by Persson et al. (2004)
and reddening estimates provided by Fernie (1995). The dashed lines 
show the edges of the instability strip. Bottom: Same as the 
top, but for the Walraven $V$,$B$-bands. 
}
\label{fig3}
\end{figure}

\subsection{Theory}
In order to provide an independent calibration of the MIC relation 
we decided to use an homogeneous set of evolutionary models for intermediate-mass 
stars with scaled-solar chemical composition. The transformation into the 
observational plane was performed by adopting a set of atmosphere models 
with the same chemical composition \citep{ca03}.  The reader interested 
in a detailed discussion concerning the input physics is referred to Pietrinferni 
et al. (2004). The adopted stellar masses range from $M=3.5 - 10.0$ $M_\odot$, 
while the metal and helium content are $Z=0.001$, $0.002$, $0.004$, 
$0.008$, $0.01$, $0.0198$, $0.03$ and $Y=0.246$, $0.248$, $0.251$, $0.256$, 
$0.259$, $0.273$, $0.288$. Fig. 3 shows the comparison between theory and 
observations for both the Johnson $V_J$,$B_J$-bands (top) and the Walraven 
$V$,$B$-bands (bottom). Individual distances were estimated using NIR mean 
magnitudes and the empirical NIR PL relations provided by Persson et al. 
(2004) and the reddening corrections provided by Fernie (1995). The dashed 
lines display the predicted first overtone 
blue edge (hotter) and the fundamental red edge (cooler) of the Cepheid 
instability strip \citep{bo05}.  

\begin{figure}
\centering
\resizebox{\hsize}{!}{\includegraphics{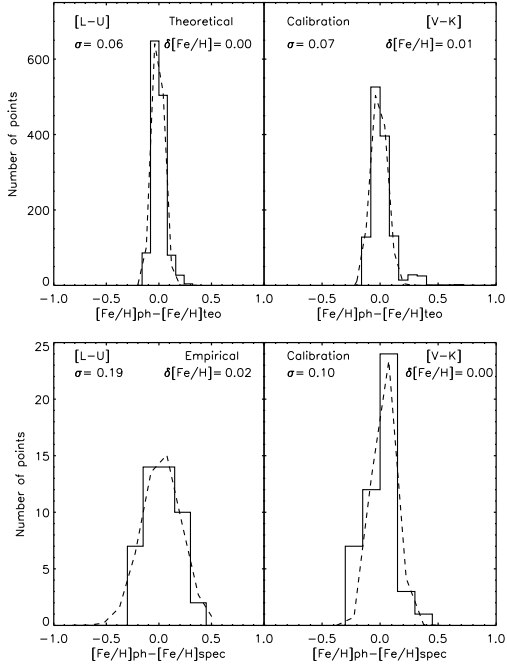}}
\caption{Top: Difference in iron abundance between theoretical MIC relations 
and spectroscopic abundances. The left panel refers to the MIC relation based on 
[L-U] color while the right one to the [V-K] color. Bottom: Same as the top, but 
for the empirical MIC relations. 
}
\label{fig4}
\end{figure}

\begin{figure}
\centering
\resizebox{\hsize}{!}{\includegraphics{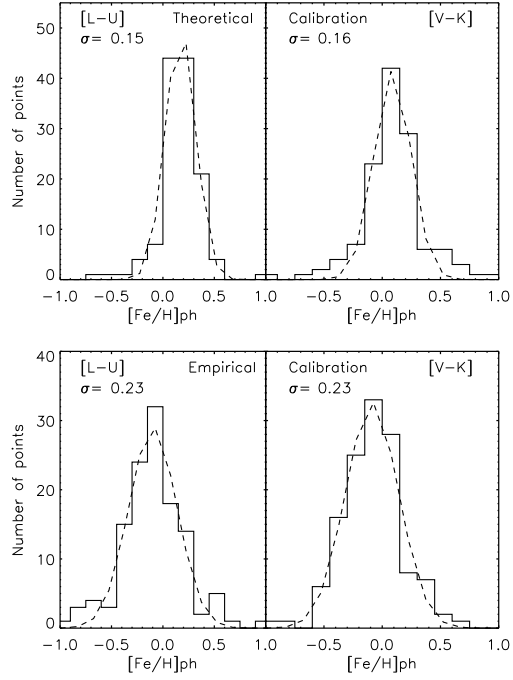}}
\caption{Metallicity distribution of Galactic Cepheids based on theoretical 
(top panels), $\sim 150$ objects, and empirical (bottom panels), $\sim 100$ 
objects. The dashed lines show the Gaussian fits, the $\sigma$ values are  
also labeled.
}
\label{fig5}
\end{figure}

\section{Metallicity indices}
In order to provide a new calibration of the MIC relation for Cepheids based 
on Walraven photometry we performed a multilinear regression fit between the 
observed (B-L) color, spectroscopic iron abundance and an independent color 
index (CI). In particular, we estimated the coefficients of the following MIC 
relation:  

{\small
\begin{equation}
(B-L)= \alpha+\beta [Fe/H]+\gamma CI_0+\delta [Fe/H] CI_0
\end{equation}
}
where the symbols have the usual meaning, and $CI_0$ indicates an unreddened 
color index. To estimate the theoretical MIC relations we selected, for each 
set of tracks at fixed metallicity (see \S 2.2), all the points located inside the 
predicted edges of the instability strip. Then we performed the multilinear 
regression fit (1) between the quoted metallicities and the predicted CIs. 
We chose two different CIs for the multilinear regression, namely
(L-U), as originally suggested by Pel \& Lub (1978), and (V-K) since optical-NIR colors 
are good temperature indicators. Moreover (L-U) is also a gravity indicator and thus 
period dependent. By adopting these two colors we found that for the calibrating 
Cepheids the differences between photometric and spectroscopic 
abundances were $\delta[Fe/H]_{(L-U)} = $ 0.03, $\sigma_{(L-U)} = $ 0.19 and 
$\delta[Fe/H]_{(V-K)} < $ 0.01, $\sigma_{(V-K)} = $ 0.11. For 
theoretical calibration $\delta[Fe/H]_{(L-U)} < $0.01, $\sigma_{(L-U)} =$ 0.08 and 
$\delta[Fe/H]_{(V-K)} = $0.01, $\sigma_{(V-K)} =$ 0.06. In order to constrain the possible occurrence of 
systematic errors introduced by reddening corrections we estimated a new set of
MIC relations, but using reddening free color indices (see \S 2.1). These relations
provide the following differences: empirical calibrations $\delta[Fe/H]_{[L-U]} = $ 0.02, 
$\sigma_{[L-U]} = $ 0.19 and $\delta[Fe/H]_{[V-K]} < $ 0.01, $\sigma_{[V-K]} = $ 0.10; 
theoretical calibrations $\delta[Fe/H]_{[L-U]} < $0.01, $\sigma_{[L-U]} =$ 0.06 and 
$\delta[Fe/H]_{[V-K]} = $0.01, $\sigma_{[V-K]} =$ 0.06.  We underline that these dispersions are 
mainly due to the standard deviation of the multilinear regression. In particular, 
we found $\sigma_{MIC([L-U])}=0.001$ and $\sigma_{MIC([V-K])}=0.006$ for theoretical 
relations and $\sigma_{MIC([L-U])}=0.14$ and $\sigma_{MIC([V-K])}=0.05$ for the 
empirical MICs. Fig. 4 shows the distribution of the difference 
between photometric estimates and spectroscopic iron measurements for 
theoretical (top panels) and empirical (bottom panels) MIC relations. Note that 
the left panels display the difference for the [L-U] colors while the right one
for the $[V-K] = (V-K) - 2.9 (B-V)$.\\
Finally, we applied the new reddening free empirical MIC relations to the non 
calibrating Cepheids ($\sim$ 100 objects) and estimated their individual iron 
abundances. The same outcome applies to the new reddening free theoretical 
MIC relations, but they have been applied to the entire sample of $\sim$ 150 
Walraven Cepheids. Data plotted in Fig. 5 show that the metallicity distributions 
based on theoretical and empirical MIC relations based on different color indices 
do agree quite well. Moreover and even more importantly, the metallicity range 
covered by the photometric abundances is, within empirical and theoretical 
uncertainties, in good agreement with spectroscopic measurements (Andrievski et al. 
2002; Yong et al. 2006; Lemasle et al. 2007)   
 
\section{Discussion}
We provide new empirical and theoretical calibrations of the Walraven 
metallicity index. We ended up with two independent sets of MIC relations 
to estimate Cepheid iron abundances. 
The comparison between iron abundance estimates based on theoretical and 
empirical MIC relations brings forward two interesting findings:\\
{\em i) Empirical Calibrations --} Cepheid iron abundances based on the empirical
calibration agree quite well with spectroscopic measurements. The difference 
is on average $0.02\pm0.02$ dex with an intrinsic dispersion $\sigma\sim 0.2$ 
dex for the relation using the [L-U] reddening free color and $0.001\pm0.002$ dex 
with $\sigma\sim0.10$ dex for the relation using the [V-K] reddening free color.\\
{\em ii) Theoretical Calibration --} Iron abundances based on the theoretical
MIC relations agree very well with spectroscopic measurements. The difference 
is on average $<0.001$ dex with an intrinsic dispersion $\sigma=0.06$ dex for 
the relation using the [L-U] reddening free color and $0.01\pm0.02$ dex with 
$\sigma=0.07$ dex for the relation using the [V-K] reddening free color. 


\bibliographystyle{aa}

\end{document}